\documentclass[aps, prl,10pt,twocolumn,superscriptaddress,nofootinbib]{revtex4-1}
\usepackage{mathrsfs, amssymb, amsmath}  
\usepackage{epsfig, cancel}
\usepackage{latexsym}
\usepackage{caption}
\usepackage{natbib, comment}
\usepackage{booktabs}
\usepackage{url}
\usepackage{eurosym}
\usepackage{dcolumn}
\usepackage{multirow}
\usepackage{color}
\usepackage{cancel}
\usepackage{soul}
\usepackage[normalem]{ulem}
\usepackage{amsfonts,amssymb,amsmath, txfonts}
\usepackage{graphicx,epsfig}
\usepackage{psfrag}
\usepackage{hyperref}
\hypersetup{colorlinks=true}
\usepackage{mathtools}
\usepackage{enumitem}
\usepackage{float}
\usepackage[dvipsnames]{xcolor}
\usepackage{xcolor}
\hypersetup{ linktoc=all,
    colorlinks, linkcolor={blue},
    citecolor={darkred}, urlcolor={darkgreen}
}

\usepackage{mathrsfs, amssymb, amsmath, mathtools} 
\usepackage{multirow}
\usepackage{amsfonts,amssymb,amsmath, txfonts}
\usepackage{graphicx,epsfig}
\usepackage{hyperref}
\usepackage{xcolor}
\usepackage{subcaption}
\usepackage{booktabs}
\graphicspath{{Images/}}
\newlength{\DefaultTabColSep}
\definecolor{rosy}{RGB}{230,235,252}
\definecolor{myframetitle}{RGB}{90,89,170}
\definecolor{myblocktitle}{RGB}{140,185,249}
\definecolor{mytitle}{RGB}{10,80,26}

\definecolor{darkgreen}{RGB}{27,130,45}
\definecolor{darkblue}{rgb}{0,0,0.3}
\definecolor{darkred}{rgb}{0.7,0,0}

\definecolor{light gray}{RGB}{220,220,220}
\definecolor{dark purple}{RGB}{108,0,217}
\definecolor{pink}{RGB}{190,20,100}
\definecolor{orang}{RGB}{193,63,0}
\definecolor{green}{RGB}{11,98,17}
\definecolor{darkpink}{RGB}{153,0,76}
\definecolor{bluegreen}{RGB}{0,102,102}
\definecolor{greenlagan}{RGB}{0,102,0}
\definecolor{redgreen}{RGB}{102,102,0}
\definecolor{Redgreen}{RGB}{153,76,0}
\definecolor{vividviolet}{rgb}{0.62, 0.0, 1.0}
\definecolor{amaranth}{rgb}{0.9, 0.17, 0.31}
\definecolor{palatinateblue}{rgb}{0.15, 0.23, 0.89}
\definecolor{brightpink}{rgb}{1.0, 0.0, 0.5}
\definecolor{cornflowerblue}{rgb}{0.39, 0.58, 0.93}
\definecolor{deepcarminepink}{rgb}{0.94, 0.19, 0.22}
\definecolor{radicalred}{rgb}{1.0, 0.21, 0.37}

\def\H0{{\text{H}\hspace*{-2.05mm}\text{H} 0\hspace*{-1.35mm}0\ }}

\def\be{\begin{equation}}
\def\ee{\end{equation}}
\def\beq{\begin{equation}}
\def\eeq{\end{equation}}
\def\bea{\begin{eqnarray}}
\def\eea{\end{eqnarray}}

\begin{document}

\title{An FWCI decomposition of Science Foundation Ireland funding}

\author{Eoin \'O Colg\'ain}
\affiliation{Atlantic Technological University, Ash Lane, Sligo F91 YW50, Ireland}

\begin{abstract}
In response to the 2008 global financial crisis, Science Foundation Ireland (SFI), now Research Ireland, pivoted to research with potential socioeconomic impact. Given that the latter can encompass higher technology readiness levels, which typically correlates with lower academic impact, it is interesting to understand how academic impact holds up in SFI funded research. Here we decompose SFI \textit{Investigator Awards} - arguably the most academic funding call - into $3,243$ constituent publications and field weighted citation impact (FWCI) values searchable in the SCOPUS database. Given that citation counts are skewed, we highlight the limitation of FWCI as a paper metric, which naively restricts one to comparisons of average FWCI ($\overline{\mathrm{FWCI}}$) in large samples. Neglecting publications with $\textrm{FWCI} < 0.1$ ($8.8\%$), SFI funded publications are well approximated by a lognormal distribution with $\mu = -0.0761^{+0.017}_{-0.0039}$ and $ \sigma = 0.933^{+0.011}_{-0.012}$ at $95 \%$ confidence level. This equates to an $\overline{\mathrm{FWCI}} = 1.433^{+0.029}_{-0.015}$ well above $\overline{\mathrm{FWCI}}=1$ internationally. Broken down by award, we correct $\overline{\mathrm{FWCI}}$ for small samples using simulations and find $\sim 67\%$ exceed \textit{median} international academic interest, thus exhibiting a positive correlation between the potential for socioeconomic impact and academic interest. 
\end{abstract}

\maketitle

\section{Introduction}
A tenet of postwar science policy is that publicly funded research trickles down to practical applications \cite{Bush:1945}. However, research may have applications decades later, making economic and societal benefits difficult to predict. Tellingly, Einstein would have struggled to foresee GPS technology as an application of general relativity \cite{Einstein:1916vd}. Succinctly put, research generating academic impact correlates \textit{positively} but \textit{weakly} with socioeconomic impact \cite{Bornmann:2012, Bornmann:2013}. 

Given the weakness of the positive correlation, can one do better? To address this question, one first appreciates what research leads to socioeconomic impact. Economically, the relationship is clearer. Starting with Griliches' seminal work \cite{Griliches:1958}, there is an appreciation that agricultural research leads to increased yields \cite{Evenson:1975, Alston:2000, Alston:2009}. Moreover, one can track the flow from academic citations through to patents \cite{Narin:1997, Ahmadpoor:2017, Li:2017, Britto:2020}. In contrast, assessing societal impact is more complicated \cite{Joly:2022, Reed:2021, Smit:2021}. Difficulties arise identifying the cause of the impact and attributing the impact to specific research, expecially with global  R\&D and innovation \cite{Martin:2007, vanderMeulenRip:2000}. 
As noted by the CoARA Transformative Research Assessment \cite{bleischwitz:2025}, \textit{``limited or non-existent funding for follow-up efforts hampers the capacity to track, understand, or enhance those impacts"}. 

In this letter, we focus on the correlation between socioeconomic and academic impact in the Irish context using Science Foundation Ireland (SFI) as a case study. Following the 2008 global financial crisis, SFI restricted funding for ``research for knowledge" to research with the potential for economic and later societal impact. Since SFI has historically not published deliverables or outputs of specific awards, assessing impact is difficult. Nevertheless, one can assume that funded projects met their socioeconomic objectives and follow up on the academic interest generated by the projects. 

The motivation here is clear, while one can identify research areas with a strong positive correlation between socioeconomic and academic impact, e. g. antiretroviral therapy for HIV/AIDS \cite{Trickey:2023}, CRISPR gene editing \cite{FajardoOrtiz:2022}, climate change \cite{Fu:2022}, etc., there is a concern that research with a high technology readiness level (TRL) based on established academic ideas may precipitate lower academic impact. \textit{A priori}, it is unclear if SFI maintained a positive correlation. A secondary goal is to establish an FWCI for a representative (median) SFI funded publication.   

In this work, we decompose SFI PI-led awards in the 5 year window 2012-2016 (budget \euro 210 million) into original research publications and their field weighted citation impact (FWCI) values using Elsevier's SCOPUS database. We show that the FWCI values are approximately lognormally distributed and that as a whole the SFI awards are internationally academically competitive. Furthermore, noting that one can only compare FWCI averages, denoted $\overline{\mathrm{FWCI}}$, in the large sample limit, we explain how one benchmarks awards with a small number of papers against median international interest. While the majority of awards $\sim 67\%$ exceed median international interest and exhibit a positive correlation between potential socioeconomic and academic impact, this leaves an appreciable minority where the value in the awards rests heavily on the socioeconomic impact generated.  

\section{Methods}

\subsection{FWCI Preliminaries}
It has been known for some time that citation counts $c$ in a field are approximately lognormally distributed \cite{Shockley:1957, Stringer:2008, Radicchi:2008, Evans:2012, Waltman:2012, Thelwall:2014, Golosovsky:2017}. In other words, one expects the logarithm of the citation count $\log c$ to be normally distributed (see \cite{Stringer:2008}). In \cite{Radicchi:2008}  Radicchi et al. made a surprising claim concerning the universality of citation counts. The claim is that one can normalise citation counts by the global average $c_0$ in the field to define a normalised count $c_f \equiv c/c_0$ that is described by a lognormal probability density function (pdf), 
\begin{equation}
\label{eq:lognormal}
    p(x) = \frac{1}{x \sigma \sqrt{2 \pi}} \exp \left( - \frac{(\ln x - \mu)^2 }{2 \sigma^2} \right), 
\end{equation}
with $x = c_f$ and parameters $\mu \in (-\infty, \infty)$, $\sigma \in (0, \infty)$. It is worth noting that since $c_f$ is normalised by the global average, further restricted to a given year and publication type, it coincides with SCOPUS' FWCI. In essence, the Radicchi et al. claim \cite{Radicchi:2008} provides a theoretical backend to the FWCI metric. 

Any lognormal distribution has a mean $e^{ \mu + \frac{1}{2} \sigma^2}$, so by normalising $c_f$ to an average of unity, one reduces the 2D parameter space $(\mu, \sigma)$ to a single parameter $\sigma^2 = - 2 \mu$. Once this is done, the surprising claim of Radicchi et al. is that each field is described by a lognormal with $\sigma^2 \approx 1.3$ \cite{Radicchi:2008}. While the universality of this statement has been questioned \cite{Albarran:2011, Evans:2012, Waltman:2012, Golosovsky:2017}, and the ideas can be further refined \cite{Radicchi:2012}, one nevertheless expects both citations and the FWCI metric to be lognormally distributed. What interests us here is that one can use the fact that FWCI values are approximately lognormally distributed with $\overline{\mathrm{FWCI}} = 1$ to benchmark the publication outputs of SFI funding. To do this, it is enough to confirm that SFI's FWCI values are lognormally distributed and that the mean exceeds $\overline{\mathrm{FWCI}} = 1$. While this may sound crude, the United Kingdom quotes a national average of $\overline{\textrm{FWCI}} = 1.54$ \cite{UKResearchBase2025}\footnote{It is unclear if a distribution is fitted or the values are simply averaged.}. A similar exercise for Ireland has yet to be performed.  

SCOPUS' FWCI metric is defined as 
\begin{equation}
    \textrm{FWCI}
=
\frac{\textrm{citations received by paper}}{
\textrm{global average for similar papers}}, 
\end{equation}
where similar papers are papers published in the same year, in the same subject category and same type of publication. The subject categories are decided by the ASJC (All Science Journal Classification) codes. If a journal makes use of multiple codes, the FWCI is calculated for each subject before averaging accordingly. It has been claimed that this classification is crude and may not suit certain fields \cite{Scelles:2025}. Any paper in SCOPUS with zero citations corresponds to $\textrm{FWCI} = 0$. However, it should be stressed that not all papers with non-zero citations in SCOPUS come with an FWCI number. We assume that this may simply be that the FWCI has yet to be calculated or that there may be combinations of year, subject and publication type where the number of publications are small and one may decide not to calculate the FWCI. We exclude publications without an FWCI value from our analysis.  

Admittedly, how the FWCI is employed in the literature is confusing. Even in the SCOPUS database, SCOPUS produces an FWCI score for individual researchers by \textit{averaging} the FWCI values for papers. While averaging works well when numbers are normally distributed and one expects to recover a number close to the median (and mode) of the distribution through the average - an approximation that gets better with larger samples - this is not the case for lognormally distributed numbers\footnote{Consider $Z$ a standard or unit normal variable with mean $\mu = 0$ and variance $\sigma^2 = 1$. Then $X = e^{\mu + \sigma Z}$
is lognormally distributed with parameters $\mu$ and $\sigma$. From here we infer that $\ln X$ is normally distributed with mean $\mu$ and standard deviation $\sigma$. Provided $X > 0$, the median value of $X$ is $e^{\mu}$, or alternatively that the average value of $\ln X$ is $\mu$, where one can recover the median value of $X$ by exponentiating the average.}. For the pdf (\ref{eq:lognormal}) the mode $e^{\mu-\sigma^2}$, median $e^{\mu}$ and mean $e^{\mu+ \frac{1}{2} \sigma^2}$ have a distinct hierarchy $\textrm{mode} < \textrm{median} < \textrm{mean}$. For this reason, any paper with $\textrm{FWCI}=1$ is in no sense an ``average" paper with median citations, but is instead in the upper percentiles where the percentile depends on $(\mu, \sigma)$. For example, for $\sigma^2 = - 2 \mu \approx 1.3$ the mean corresponds to the $72^{\textrm{nd}}$ percentile placing it well above ``average" ($50^\textrm{th}$ percentile) in the usual sense. 

This makes the FWCI misleading, since it is marketed as a metric that one can use at the level of an individual paper; each paper gets its own FWCI score. In general, if citation counts $c$ are lognormally distributed, so that $\log c$ is normally distributed , one can employ any normalisation one chooses\footnote{Note that the normalising factor $c_0$ simply shifts the normal distribution by a factor of $\log c_0$.}. Adopting the median citation count as the normalising factor, one would find that any paper with $\mathrm{FWCI}>1$ is in the upper $50 \%$ of papers. This is a much better metric to apply to an individual paper. 

Instead, if one normalises with respect to the average, then comparisons only make sense at the level of a lognormal distribution and this requires a relatively large sample to be sure that the distribution is lognormal. For this reason, University College Cork \cite{UCC} provides guidance that the FWCI be only applied to institutions with 1000's of publications. As we shall show later with simulations, samples of $n=400$ may allow comparisons to be made with as little as $1 \%$ approximation for $\sigma^2 = - 2 \mu \in [1, 1.8]$ \cite{Radicchi:2008}. Given that it only makes sense to compare $\overline{\mathrm{FWCI}}$ with large samples, one can assess small samples by benchmarking them against simulations with lognormals with $\sigma^2 = - 2 \mu \in [1, 1.8]$. We will illustrate how this can be done in due course.   

The FWCI has the usual blind spots that citation counts possess. While the FWCI corrects for subject, publication type and year, it does not correct for self-citations or researchers working in larger collaborations to enhance citations \cite{Wuchty:2007, shen:2021}. However, if self-citations are prevalent in a subject area, they get averaged. A further drawback of the FWCI is its short shelf-life. The FWCI tracks changes in citations for the original year and three subsequent years, so it only tracks academic impact dynamically in the short-term and beyond 4 years maximum the FWCI value of a paper is fixed. Note, the rate at which a paper's citations increase, peak and then decline over time depends on the field \cite{galiani:2017, Gou:2022}, but this is addressed in the field weighting. Evidently, for fields where the citations increase more slowly, the FWCI becomes noisier because increases in citation count are more dramatic with lower citations. Undeniably, the short time window precludes ``sleeping beauty" papers that generate academic interest much later. This may be a shortcoming, but only $\sim 0.01 \%$ of papers are estimated to be in this category \cite{vanRaan:2004, Ho:2017, Teixeira:2017}; from our sample of 3000 odd papers, one expects less than 1 publication. 

Finally, the FWCI tracks research interest and not research quality. That being said, the UK's Research Excellence Framework (REF) \cite{Thelwall:2023} shows a positive correlation between FWCI/citations and quality across all subjects with a correlation that is stronger in STEM subjects than humanities. It is worth also emphasising that one cannot generate academic impact without first generating the academic interest tracked by the FWCI.

In summary, to a degree of approximation one expects citation counts  $c$ to be lognormally distributed \cite{Shockley:1957, Stringer:2008, Radicchi:2008, Evans:2012, Waltman:2012, Thelwall:2014, Golosovsky:2017}. As a consistency check, one can study $\log c$ and check if the values are normally distributed. If one divides the citation count in a given field by the global average in a given year for a given publication type, one recovers SCOPUS' FWCI metric, which one expects to be lognormally distributed with mean $\overline{\mathrm{FWCI}} = e^{\mu + \frac{1}{2} \sigma^2} = 1$ by construction. As the United Kingdom have done, this provides a means to benchmark research output. Here we apply it to the SFI \textit{Investigator Awards} by checking if the $\overline{\mathrm{FWCI}}$ exceeds the global average. Furthermore, for small samples we benchmark $\overline{\mathrm{FWCI}}$ against the median from simulations.

\subsection{SFI Data}
To compile our dataset we extract projects YY/IA/XXXX from SFI Open Data \cite{sfi:opendata}, where YY $\in \{12, 13, 14, 15, 16\}$ denotes the evaluation year of the proposal and XXXX is the award code. The awards are usually announced in the annual report \cite{sfi:reports} of the subsequent year\footnote{There are a few 2016 proposals announced in the 2018 annual report.}. We note that the budgets in SFI annual reports need not agree with the budgets in SFI Open Data, since after funding has been allocated, there are supplements, e. g. maternity leave, and funding can be de-committed. Here ``IA"  is the identifier for \textit{Investigator Awards}. It is worth noting that these awards are typically made to senior academics/professors with arguably the strongest grasp of academic interest, so within SFI funding, one may expect the awards to be academically competitive despite SFI's socioeconomic remit.  

Given a project code, one can either search by the code directly in Google Scholar or by the \textit{funding number} field in Elsevier's SCOPUS. Given our interest in the FWCI, we restrict our interest to SCOPUS, but use Google Scholar for crosschecks. We summarise the yearly budgets, the number of awards and publications in Table \ref{tab:budgets}, where it should be stressed that we only include publications with an FWCI value searchable in SCOPUS. We search by both YY/IA/XXXX and SFI/YY/IA/XXXX, which is commonly used in lieu. We exclude publications classified as \textit{review}, \textit{editorial}, \textit{short survey} and \textit{book chapter} on the grounds that they need not constitute original research and are often subject to lower peer-review scrutiny. Ultimately, our goal is to gain an insight into original research funded by SFI. Although conference papers are a grey area, and they need not be peer-reviewed, in the fields of computer science and engineering conference papers are peer-reviewed and may be more important than articles. In short, we focus on publications classified as \textit{article}, \textit{conference paper}, \textit{letter} and \textit{note}, but the latter two infrequently arise. We extracted the data from SCOPUS in early January 2026. SCOPUS only tracks FWCI numbers for publications for a maximum of 4 years, so this means that publications in the awards through to 2021 no longer have dynamical FWCI numbers.  

The data comes with some caveats. First, the SCOPUS coverage of the literature may be incomplete \cite{MartinMartin:2018a, MartinMartin:2018b}. Secondly, SCOPUS occasionally misclassifies reviews as original research articles. Thirdly, one can find examples where the SFI project code is correctly recorded in the acknowledgement text, but the code has not propagated to the \textit{funding number} field in SCOPUS metadata, so does not appear in searches. However, it should be stressed that we do not need an exhaustive list of papers for each project, it is enough to have a list of papers and their FWCI numbers to get an insight into the academic interest generated by an award. Our SCOPUS funding number searches return on average 20 publications per project, which is sufficient to build up a picture of the potential academic impact of the project. 

\begin{figure}[htb]
   \centering
\includegraphics[width=90mm]{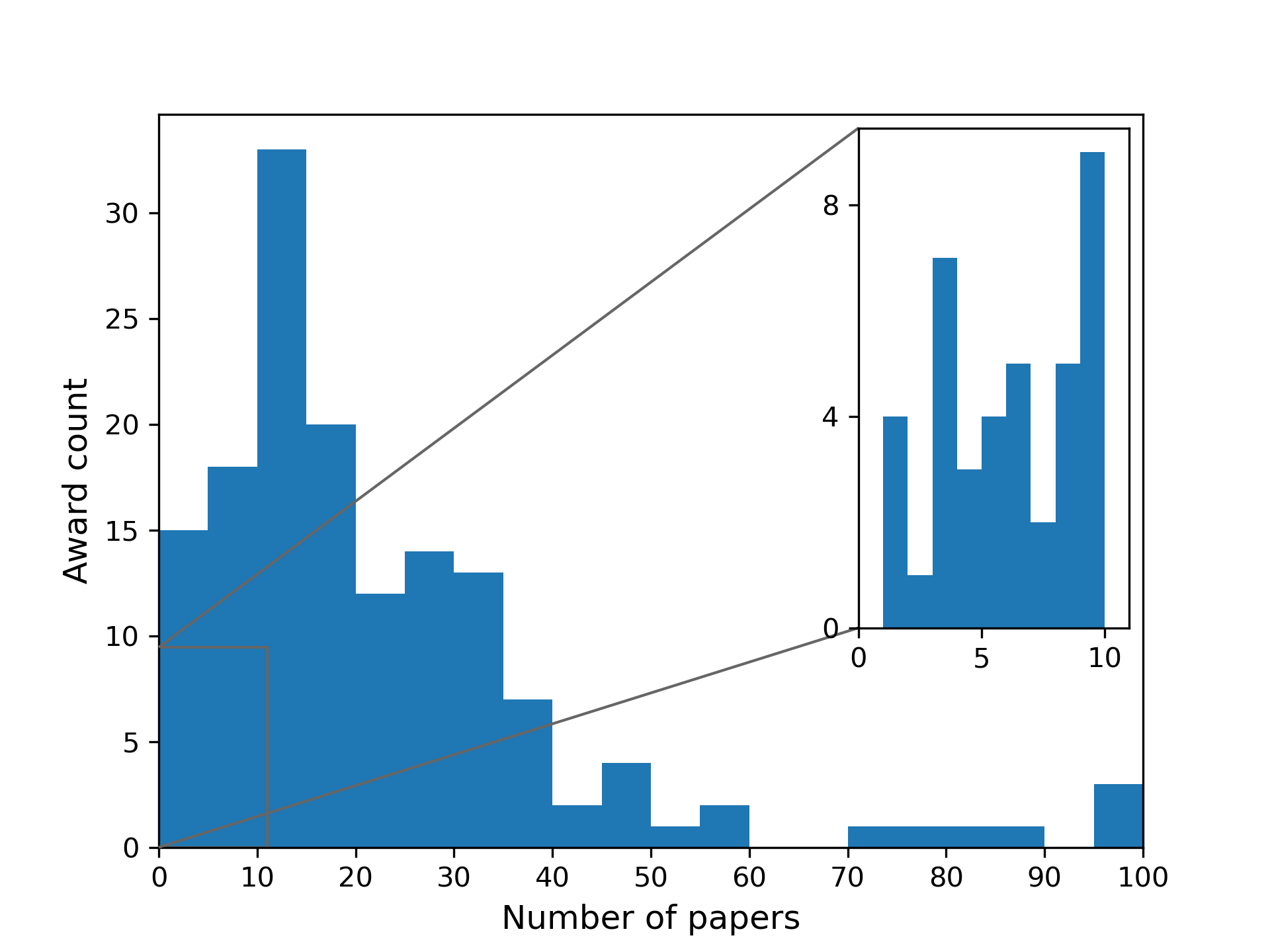}
\caption{A breakdown of the 148 awards by number of papers traceable using \textit{funding number} metadata in SCOPUS. We removed awards with no original publications and provide a zoom-in for awards with less than 10 papers.}
\label{fig:papers} 
\end{figure}

In Fig. \ref{fig:papers} we document the number of publications we identify for each award, where we find the most likely number of publications to be in the 10-15 window. The differences can largely be explained by subject area. From Table \ref{tab:budgets} the average budget of each award is \euro $1.42$ million and the average cost of an original research publication is \euro$65,000$. These figures are indicative. Given that the number of papers varies across the awards in Fig. \ref{fig:papers}, it should be clear that the cost of a paper varies. We remind the reader again that owing to the caveats concerning incompleteness of our dataset, \euro$65,000$ represents an upper bound on the average cost of a paper. 

%
%

\begin{table}

\begin{tabular}{cccc}
\rule{0pt}{3ex}    \textbf{Year} & \textbf{Awards} & \textbf{Publications} & \textbf{Budget (\euro)} \\
\hline 
\hline
\rule{0pt}{3ex}    2012 & 31 & 591 & 41,974,909 \\
\rule{0pt}{3ex}    2013 & 39 & 960 & 49,382,219 \\
\rule{0pt}{3ex}    2014 & 23 & 466 & 27,333,184 \\
\rule{0pt}{3ex}    2015 & 23 & 470 & 38,239,664 \\
\rule{0pt}{3ex}    2016 & 32 & 756 & 53,151,712 \\
\hline
\hline
\rule{0pt}{3ex} \textbf{Total} & 148 & 3,243 & 210,081,688
    
\end{tabular}
\caption{A breakdown of SFI's Investigator Awards (IA) into number of awards, number of publications with FWCI and total budget by the evaluation year. We comment on omitted awards in the text.}
\label{tab:budgets}
\end{table}

SFI's \textit{Investigator Awards} include 11 additional awards that do not appear in Table \ref{tab:budgets}. We identified 4 awards (12/IA/1570; 12/IA/1547; 13/IA/1806; 16/IA/4628) without a record of an original research publication in either SCOPUS or Google Scholar. It is plausible that publications not referencing the code exist or simply that the deliverables did not include academic publications. Separately, it has been shown that $2.4 \%$ of National Institute of Health (NIH) grants, including research on humans and clinical trials that typically delay publications, may not produce a publication within 5 years of the initiation of the grant \cite{Riley:2020}. Here the percentage is comparable but we are well beyond the 5 year window. To back up the possibility that award numbers need not be properly acknowledged, we find isolated cases where the project numbers are incorrectly documented, e. g. \cite{oreilly:2021} is transferred between two SFI awards, or that typos in project numbers lead to phantom awards downstream in the SCOPUS database\footnote{The vast majority of publications of the award 14/IA/2508 appear as 14/1A/2508 in SCOPUS. In addition, 14/IA/2884 appears in SCOPUS, but the closest official record in SFI Open Data \cite{sfi:opendata} is 14/IA/2284 with two publications documenting both numbers. Where the typo is obvious, we included the paper in our analysis.}. Finally, we exclude 7 awards (15/IA/2864; 15/IA/2881; 15/IA/3028; 15/IA/3058; 15/IA/3136; 15/IA/3152; 15/IA/3160) co-funded with the Ministry of the Economy of Northern Ireland \cite{SFI-DfE} on the grounds that the co-PIs share a common award number, yet papers may be exclusively written with a Northern Ireland affiliation.

\section{Results}

\begin{figure}[htb]
   \centering
\includegraphics[width=90mm]{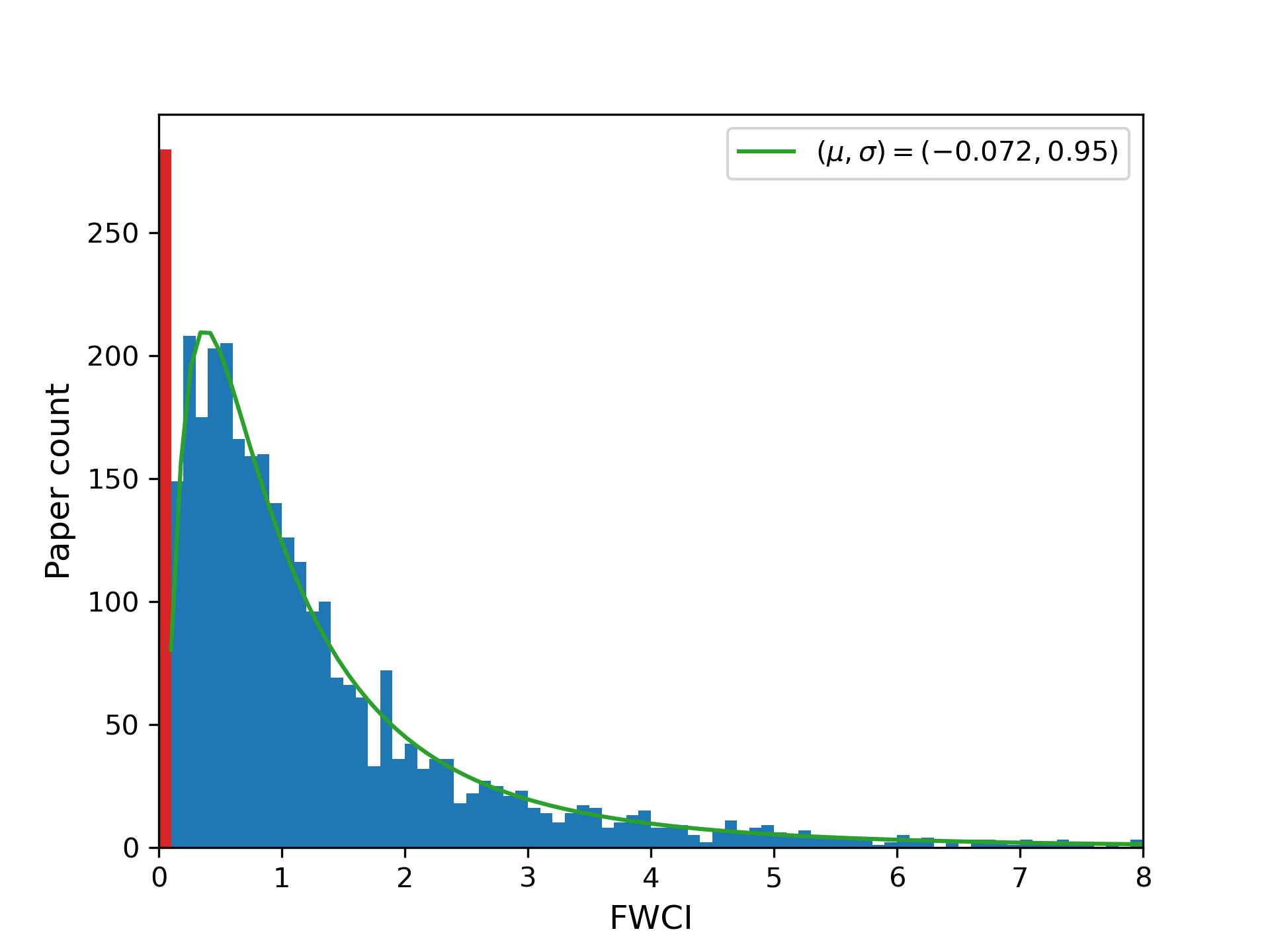}
\caption{A histogram of FWCI values from 3,243 publications from SFI awards with $\mathrm{FWCI} < 0.1$ publications highlighted in red. The green curve is the best fit lognormal to the blue bins.}
\label{fig:lognormal} 
\end{figure}

We begin with the full sample of 3,243 publications without decomposing into the 148 awards. In Fig. \ref{fig:lognormal} we show the count of FWCI numbers in bins of $\Delta \textrm{FWCI} = 0.1$. Following Evans et al. \cite{Evans:2012}, who removed it from their analysis, we highlight the $\textrm{FWCI} < 0.1$ bin in red. These 284 publications correspond to $8.8 \%$ of the full sample and include both uncited publications, $\textrm{FWCI} = 0$, and low FWCI papers, $0.04 \leq \textrm{FWCI} < 0.1$. 

\begin{figure}[htb]
   \centering
\includegraphics[width=90mm]{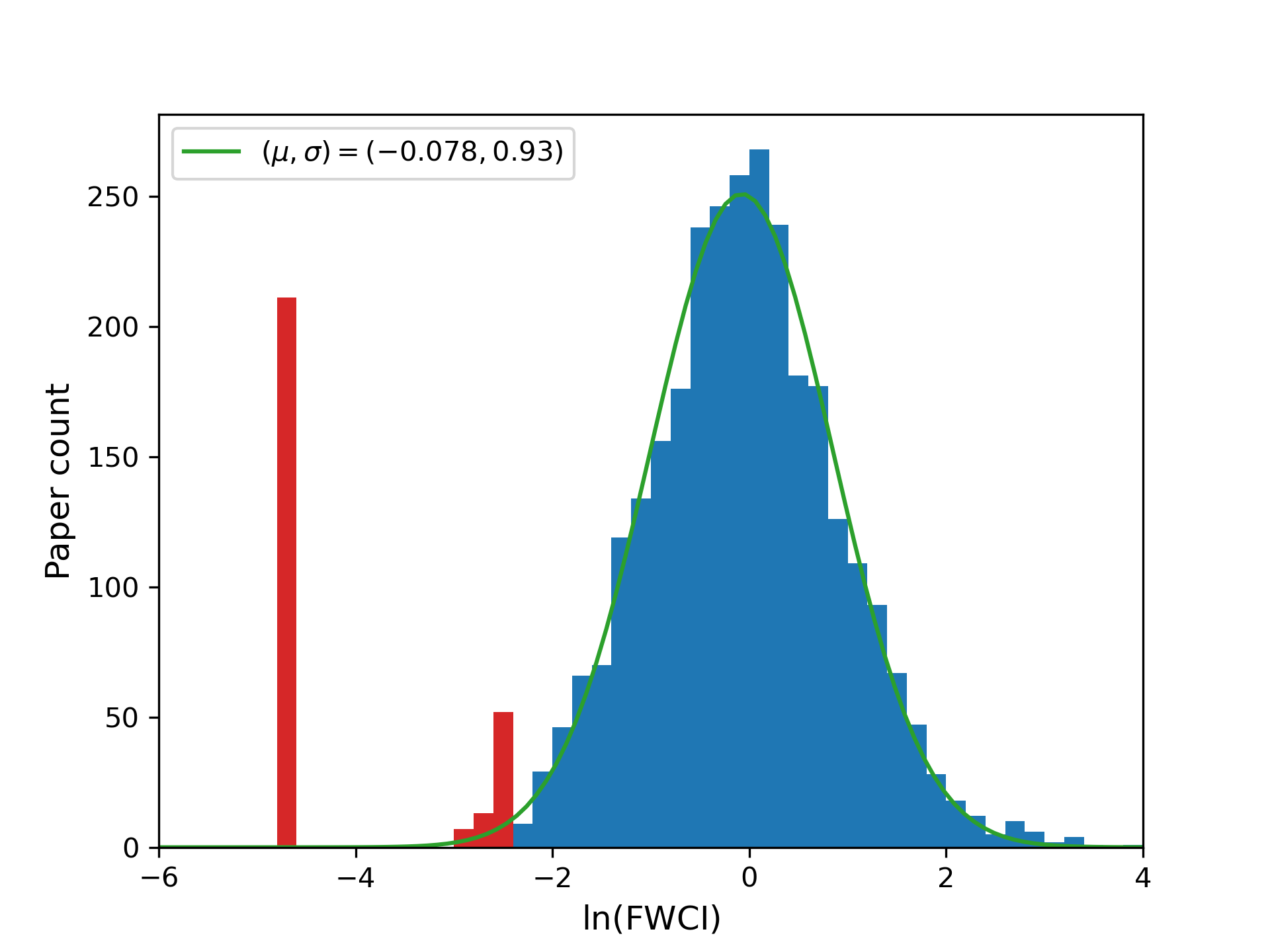}
\caption{A histogram of the natural logarithm of FWCI values from Fig. \ref{fig:lognormal}. $\mathrm{FWCI} < 0.1$ publications are highlighted in red where we have shifted $\mathrm{FWCI} =0 \rightarrow 0.01$ papers for visual purposes. The green curve is the best fit normal to the blue bins.}
\label{fig:normal} 
\end{figure}

In Fig. \ref{fig:normal} we take the natural logarithm of the FWCI values in Fig. \ref{fig:lognormal} and confirm that two exceptions aside the natural logarithm $\ln (\mathrm{FWCI})$ is approximately normally distributed in bins of $\Delta \ln (\mathrm{FWCI}) = 0.2$. Once again, the red bins correspond to $\textrm{FWCI} < 0.1$, where we displaced the uncited papers $\textrm{FWCI} = 0 \rightarrow 0.01$ to a small value below the smallest finite value for visual purposes, otherwise they would map to $\ln(\mathrm{FWCI}) = -\infty$. These uncited papers explain the peak at $\ln(0.01) \approx -4.6$ in Fig \ref{fig:normal} that is clearly displaced from the normal. This provides visual confirmation that uncited papers cannot be normally distributed, and thus cannot be lognormally distributed, and should be removed from the analysis. The remaining $0.04 \leq \textrm{FWCI} < 0.1$ papers constitute the secondary bump at the lower tail end of the normal. These also do not visually conform to the normal and this justifies the choice made in ref. \cite{Evans:2012} to remove them. 

Once the red bins are removed, the green curves constitute the best fit lognormal and normal in Fig. \ref{fig:lognormal} and Fig. \ref{fig:normal}, respectively. We only fit a normal to $\ln (\mathrm{FWCI})$ as a consistency check that we get more or less the same best fit $(\mu, \sigma)$ values. Since we have not normalised the paper count, we fit the function
\begin{equation}
 p ( x) = \frac{A}{x} \exp \left( - \frac{(\ln x - \mu)^2 }{2 \sigma^2} \right), 
\end{equation}
where $x = \mathrm{FWCI}$ and $A$ is an auxiliary parameter accounting for the scale. It should be noted that the best fit values of $(\mu, \sigma)$ depend on the number of bins when one fits either a lognormal or normal and this explains the slight differences in $(\mu, \sigma)$ quoted in the plots. 

In Fig. \ref{fig:lognormal} we only fit the data in the range $\textrm{FWCI} \in (0, 8)$ but we have checked that relaxing this to $\textrm{FWCI} \in (0, 50)$ while maintaining $\Delta \textrm{FWCI} = 0.1$ to accommodate our largest FWCI value $\textrm{FWCI} = 49.27$ returns the same best fit values. Nevertheless, it should be clear from either histogram that the best fit $(\mu, \sigma)$ values depend on the width of the bins. We address this as follows. We restrict our attention to $\textrm{FWCI} \in (0,8)$ but allow the number of bins to vary randomly and uniformly in the range $[20, 800]$ 10,000 times. This leads to 10,000 histograms and for each histogram we fit $(A, \mu, \sigma)$ to identify a distribution of the parameters of interest, $(\mu, \sigma)$. From this distribution, we isolate the median ($50^{\textrm{th}}$ percentile) as the central value and use the $2.5$ and $97.5$ percentiles to define the limits of $95 \%$ confidence intervals. The resulting values for the SFI awards are 
\begin{equation}
\label{eq:best_fit}
    \mu = -0.0761^{+0.017}_{-0.0039}, \quad \sigma = 0.933^{+0.011}_{-0.012}. 
\end{equation}
At this point we recall that if FWCI numbers in a field are described by a lognormal as claimed by Radicchi et al. \cite{Radicchi:2008} with $\mu = - \frac{1}{2} \sigma^2 = -0.65$ where we have inserted $\sigma^2 = 1.3$, then $\mu > -0.65$ confirms that the papers corresponding to the blue bins in Fig. \ref{fig:lognormal} outperform the international average. Even if one allows for the lowest value in \cite{Radicchi:2008}, $\sigma^2 = 1$, one still confirms $\mu > -0.5$. In effect, this confirms that taken as a whole, the papers in SFI awards are the result of an academically competitive process. 

From exponentiating $\mu$ in (\ref{eq:best_fit}), one finds a median for an SFI publication: 
\begin{equation}
    \label{eq:median}
    e^{\mu} = 0.9267^{+0.016}_{-0.0035}, 
\end{equation}
where the antisymmetric errors are due to the skewness of the distribution. Thus, $\mathrm{FWCI} \sim 0.93$ is a representative value for an SFI publication. From the central values in (\ref{eq:best_fit}), one can also infer that the $95 \%$ confidence interval for SFI publications is $0.03 \lesssim \mathrm{FWCI} \lesssim 4.53$. We remind the reader again that these numbers do not incorporate uncited and lower FWCI papers.   

Noting that one can extract a mean 
\begin{equation}
\label{eq:confidence}
    e^{\mu + \frac{1}{2} \sigma^2} = 1.433^{+0.029}_{-0.015}
\end{equation}
at $95 \%$ confidence level from (\ref{eq:best_fit}), this similarly confirms that the mean of the SFI lognormal exceeds the global average $\overline{\textrm{FWCI}} =1$. It is interesting to compare this to the mean one would get without fitting a lognormal, $\overline{\textrm{FWCI}} = 1.58$. Given that this number falls outside our $95 \%$ confidence interval (\ref{eq:confidence}), it is reasonable to assume that the lognormal is only an approximation and the tails may be heavier. If one further includes the $\textrm{FWCI} < 0.1$ papers, this reduces to $\overline{\textrm{FWCI}} = 1.43$. We observe that these numbers are comparable to the United Kingdom's national average $\overline{\mathrm{FWCI}} = 1.54$, so SFI's $\overline{\mathrm{FWCI}}$ value transplanted to the United Kingdom may not be competitive. This underscores the importance of understanding the Irish national baseline.   

\begin{figure}[htb]
   \centering
\includegraphics[width=90mm]{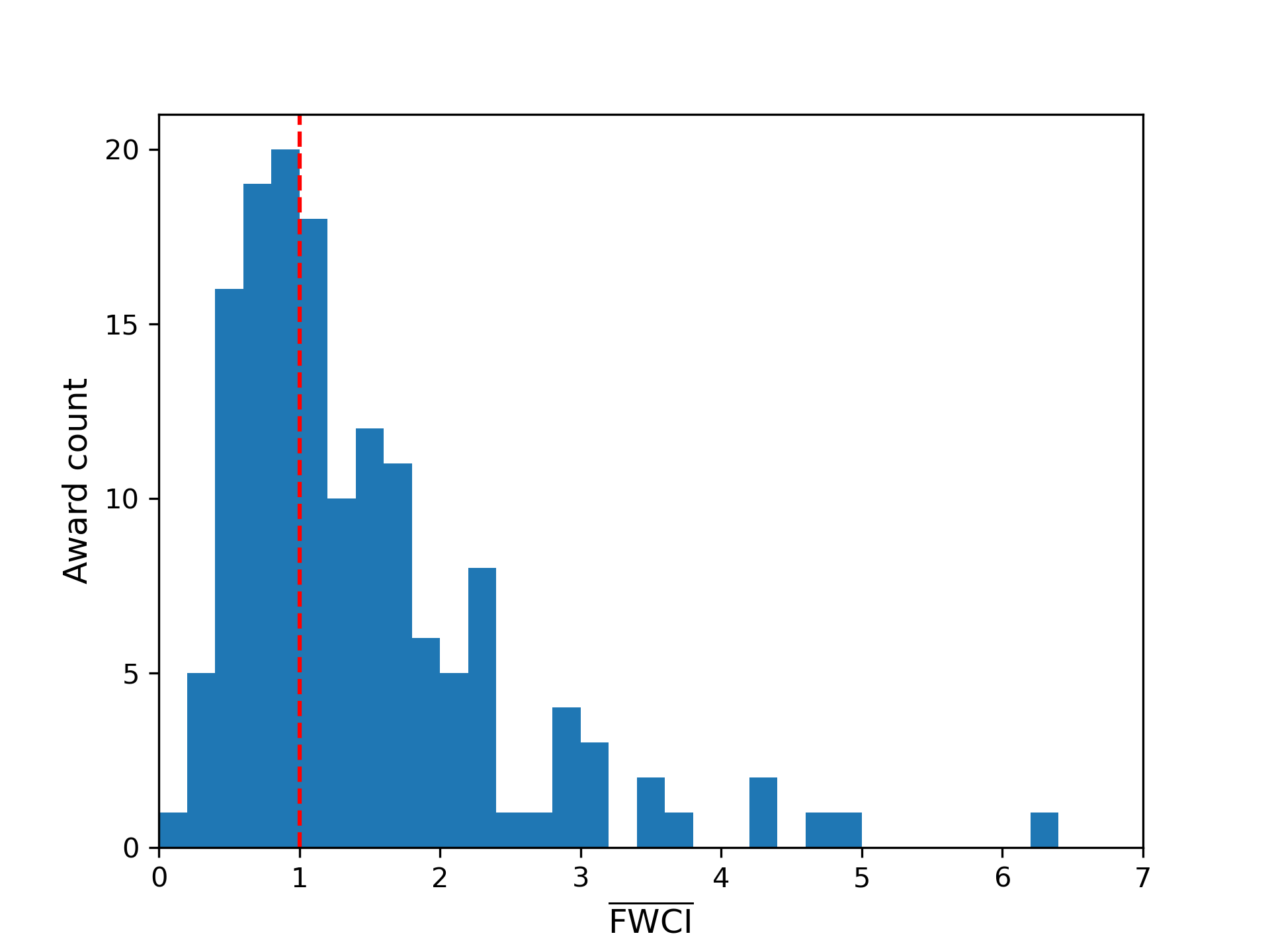}
\caption{A breakdown of the average $\textrm{FWCI}$ for each of the 148 awards in Table I. 87 from 148 awards  ($59 \%$) have  $\overline{\mathrm{FWCI}} \geq 1$.}
\label{fig:average} 
\end{figure}

Broken down by project we get a better appreciation for the strengths and weaknesses of the awards in generating academic interest. In Fig. \ref{fig:average} we show that $59 \%$ of the awards possess $\overline{\mathrm{FWCI}} \geq 1$. Naively, this says that $59 \%$ of the awards have safely led to academic interest at the level of global average citations or better. Bearing in mind that the original paper count in awards can be low - see inset from Fig. \ref{fig:papers} - demanding $\overline{\mathrm{FWCI}} \geq 1$ sets a high bar. We remind the reader again that a single paper $\mathrm{FWCI}=1$ places the paper at approximately the $72^{\mathrm{nd}}$ percentile, thus well above the median. Ultimately, as advocated in the UCC report \cite{UCC}, comparisons based on $\overline{\mathrm{FWCI}}$ are only reliable in the large sample limit.    

What remains to be done now is to make allowances for the fact that awards have a much smaller number of papers. There is an easy way to do this by pegging the awards to the median and not the average. We begin by assuming that FWCI scores in a field are lognormally distributed with $\mu = - \frac{1}{2} \sigma^2$ and  $\sigma^2 \approx 1.3$ \cite{Radicchi:2008}. For a single paper this corresponds to a median ($50^{\mathrm{th}}$ percentile) FWCI score of approximately $e^{-0.65} = 0.52$. Thus, an award with a single paper with $\mathrm{FWCI} \gtrsim 0.52$ could be considered above average by inhabiting the upper $50\%$ of comparable papers in the same field, same year and with the same publication type. Note also that as $\sigma$ increases, $\mu$ decreases, so for $\sigma^2=1$, the lowest number quoted by Radicchi et al. \cite{Radicchi:2008}, the median shifts up to $e^{-0.5} = 0.61$. For this reason, our assumption that FWCI numbers are lognormally distributed with $\mu = -\frac{1}{2} \sigma^2 = - 0.65$ is a relatively middle of the road assumption. 

\begin{figure}[htb]
   \centering
\includegraphics[width=90mm]{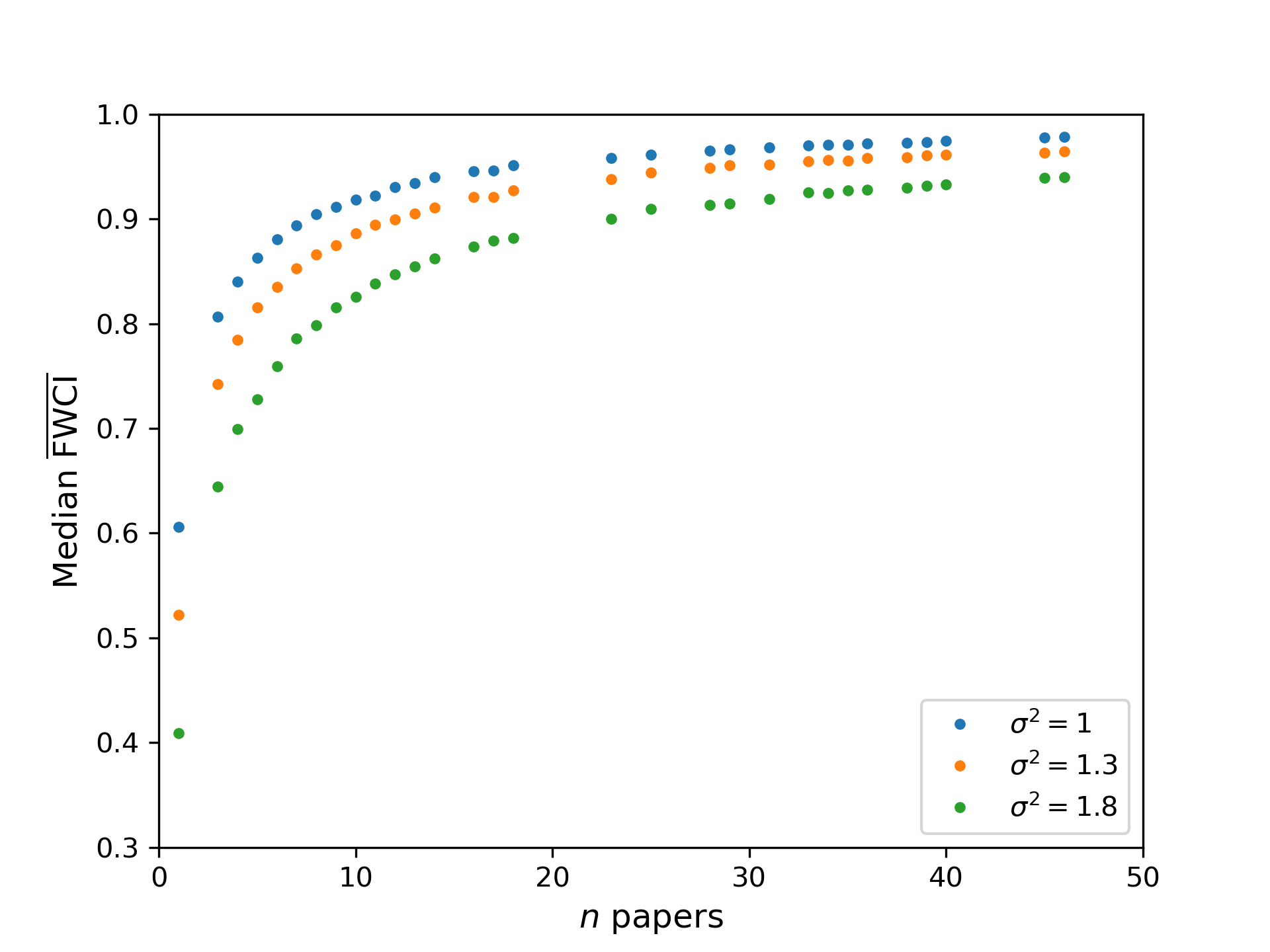}
\caption{The median $\overline{\mathrm{FWCI}}$ from the $\overline{\mathrm{FWCI}}$ of $n$ papers drawn randomly from lognormal distributions with $\sigma^2 = - 2 \mu$ across 100,000 simulations. For $n=1$ one recovers the median of a lognormal $e^{-\frac{1}{2} \sigma^2}$.} 
\label{fig:corrected_fwci} 
\end{figure}

From here, we run simulations to proceed. For each of the 61 awards with $\overline{\mathrm{FWCI}} < 1$ we identify the number of papers in the award $n$ and generate 100,000 copies of the award with $n$ papers in a lognormal distribution with $\mu = -\frac{1}{2} \sigma^2$ with $\sigma^2 \in \{1, 1.3, 1.8\}$, the range of values from Radicchi et al. \cite{Radicchi:2008}. For each copy of the award we calculate an average $\overline{\mathrm{FWCI}}$ and finally we identify the median of these 100,000 $\overline{\mathrm{FWCI}}$ values. The result of this exercise is shown in Fig. \ref{fig:corrected_fwci} where the number of papers $n$ is taken from the 61 awards. Evidently, at $n = 46$ the median $\overline{\mathrm{FWCI}}$ is within $10\%$ of $\overline{\mathrm{FWCI}} = 1$ and the median increases to $\overline{\mathrm{FWCI}} = 1$ as $n$ increases. For $n=400$ we find that median $\overline{\mathrm{FWCI}}$ is within $1\%$ of $\overline{\mathrm{FWCI}} = 1$ even for $\sigma^2 = 1.8$, so one is clearly close to the large sample limit. 

This is the benchmark against which we compare the 61 awards with $\overline{\mathrm{FWCI}} < 1$ to ascertain whether the observed award average exceeds the median from our simulations based on the assumption that the underlying FWCI is lognormally distributed \cite{Radicchi:2008}. We find from the 61 awards with $\overline{\mathrm{FWCI}} < 1$ that a further 9 ($\sigma^2=1$), 12 ($\sigma^2=1.3$) or 15 ($\sigma^2=1.8$) awards may be considered to be above international median academic interest when correcting for the small number of papers in awards. Evidently, as $\sigma^2$ increases, the international median $\overline{\mathrm{FWCI}}$ decreases making the SFI awards more competitive.   

In summary, at the level of awards, we find that 96 to 102 awards (central value 99) from 148 awards, corresponding to $65 \%-69\%$ (central value $67\%$) may be considered to have attracted international academic interest that places them in the top $50 \%$. It is worth stressing to get this result we have assumed that internationally FWCI values are lognormally distributed with $\sigma^2 \in [1, 1.8]$.

\section{Discussion}
The funding landscape in Ireland requires all research proposals to be viewed through the prism of socioeconomic impact. Traditionally, the correlation between academic and socioeconomic impact is positive but weak \cite{Bornmann:2012, Bornmann:2013}, so it is interesting to see how academic interest holds up when one attempts to strengthen the correlation as SFI has done. \textit{A priori,} one can easily increase the prospect of socioeconomic impact by focusing on local/regional problems and high TRL research, but this risks undermining academic impact. Ultimately, assessing research impact, both academic and socioeconomic impact, requires rigorous follow-up \cite{bleischwitz:2025}.   

In this letter we studied the academic interest of SFI funded \textit{Investigator Awards} through SCOPUS' FWCI metric. Since SFI does not document the outputs from research, this involved a reconstruction of the awards. Here, one cannot have academic impact without academic interest. The FWCI metric is attractive in that citation counts in fields are lognormally distributed to good approximation \cite{Radicchi:2008, Stringer:2008}, allowing one to make comparison between fields. However, the FWCI is normalised with respect to a global average, and due to the skewness of citation counts this biases it away from the median and towards higher percentiles; a globally ``average'' paper with $\mathrm{FWCI}=1$ may correspond to the $70^{\textrm{th}}$ percentile. For this reason, FWCI as a paper metric is misleading and it is more a metric for comparing lognormal distributions from large samples. Nevertheless, with large samples of papers and their FWCI scores, one can use $\overline{\mathrm{FWCI}}$ to benchmark publications. Moreover, through simulations and the assumption that FWCI scores are lognormally distributed in fields with $\mu = - \frac{1}{2} \sigma^2$ and $\sigma^2 \sim 1.3$ \cite{Radicchi:2008}, one can correct $\overline{\mathrm{FWCI}}$ to benchmark smaller samples against the median. 

Here, we have applied both methods to SFI \textit{Investigator Awards} finding that when treated as a large sample, the publications are internationally competitive, and when broken down into awards, $65\%$ to $69\%$ of the awards appear to exceed median international academic interest. It is unclear if the remaining $\sim 30 \%$ awards were targeting socioeconomic impact and not academic impact from the outset or simply projects with less marketable results. Regardless, as $\overline{\mathrm{FWCI}}$ decreases, the value in the awards rests more on socioeconomic impact. Given that PI-led awards are expected to be some of the most academically competitive, our findings may be in line with expectations. It would be interesting to repeat the exercise with SFI/Research Ireland funding where there may be less of an academic focus.   

In general the methods can be applied at a national level, as the United Kingdom government has done \cite{UKResearchBase2025}. This would allow one to understand how much of the SFI result is due to baseline academic interest generated by Irish papers and how much is due to the SFI selection process. At this juncture, this is unclear. Furthermore, one can apply the methods to benchmark universities, which will be interesting in the Irish context as the fledgling Technological Universities pivot from teaching to focus on research with regional and socioeconomic impact, where it will be interesting to see if the research also generates academic impact.

\section*{Acknowledgements}
We thank Hirushan Sajindra and Shahin Sheikh-Jabbari for discussions on related topics. 

\bibliography{refs}

\end{document}